\documentclass[12pt]{article}
\def\be{\begin{equation}}

\def\ee{\end{equation}}
\def\bea{\begin{eqnarray}}
\def\eea{\end{eqnarray}}
\title{$O(D)$ invariant tachyon condensates in the $\frac{1}{D}$ expansion}
\author{G. Grignani$^1$, M. Laidlaw$^2$, \\
M. Orselli$^4$ and G. W. Semenoff$^2$\\ ~~ \\ 1.)Dipartimento di
Fisica and Sezione INFN, Universit\`a di Perugia, \\Via A. Pascoli
I-06123, Perugia, Italia\\ ~~ \\ 2.)Department of Physics and Astronomy,
University of British Columbia, \\ Vancouver, British Columbia, Canada
V6T 1Z1 \\ ~~ \\
3.) Dipartimento di Fisica and
Gruppo Collegato INFN, Universit\`a di Parma,\\ Parco Area della Scienze 7A, I-43100
Parma, Italia}

\begin{document}           

\maketitle                 
\hspace{4.5in}{UPRF-2002-05}

\abstract{ We consider the problem of condensation of open string
tachyon fields  which have an $O(D)$ symmetric profile.  This problem
is described by a boundary conformal field theory with $D$ scalar
fields on a disc perturbed by relevant boundary operators with $O(D)$
symmetry.  The model is exactly solvable in the large $D$ limit and we
analyze its $1/D$ expansion. We find that this expansion is only
consistent for tachyon fields which are polynomials.  In that case, we
show that the theory is renormalized by normal ordering the
interaction.  The beta-function for the tachyon field is the linear
wave operator.   We derive an expression for the tachyon potential and
compare with other known expressions.   In particular, our technique
gives  the exact potential for the quadratic tachyon profile.  It can
be used to correct the  action which has been derived in that case
iteratively  in derivatives of the tachyon field.}

\newpage

A classic problem in string theory is to understand how the background
space-time on which the string propagates arises in a self-consistent
way.  For open strings, there are two main approaches to this problem,
cubic string field theory  \cite{Witten:1985cc} and background
independent string field theory \cite{Witten:1992qy,Shatashvili:1993kk}.

The latter approach is defined as a problem in boundary conformal
field theory. One begins with the partition function of open-string
theory where the world-sheet is a disc.  The strings in the bulk are
considered to be on-shell and a boundary interaction with arbitrary
operators is added.  The configuration space of open string field
theory is then taken to be the space of all possible boundary
operators modulo gauge symmetries and the possibility of field
re-definition.  Renormalization fixed points, which correspond to
conformal field theories, are solutions of classical equations of
motion and should be viewed as the solutions of classical string field
theory.

Despite many problems which are both technical and matters of
principle, background independent string field theory has been useful
for finding the classical tachyon potential energy functional and the
leading derivative terms in the tachyon effective action
\cite{Gerasimov:2000zp,Kutasov:2000qp,Tseytlin:2000mt}
\footnote{For earlier works on tachyon condensation see 
\cite{Bardakci:ui,Bardakci:1974vs,Bardakci:1975ux,Bardakci:1977an}.}.
Boundary
field theories which can be used to study tachyons in the background
independent string field theory framework are the subject of the
present paper.

The existence of a tachyon in the bosonic string theory indicates that
the 26-dimensional Minkowski space background about which the string
is quantized is unstable. An unstable state should decay to something
and the nature of both the decay process and the endpoint of the decay
are interesting questions.  Recently, some understanding of this
process has been achieved for the open bosonic string.  The key idea
is that of Sen~\cite{Sen:1999mh}.   The open bosonic string tachyon
reflects the instability of the D-25 brane.  This unstable D-brane
should decay by condensation of the open string tachyon field.  The
energy per unit volume released in the decay should be the D-25 brane
tension and the end-point of the decay is the closed string vacuum
\cite{Sen:1999mh,Elitzur:1998va,Harvey:1999gq,Kutasov:2000qp}. There
are also intermediate unstable states which are the D-branes of all
dimensions between zero and 25.
To study  background independent string field theory, 
consider the partition function
\begin{equation}
Z=\int [d X^j(\sigma,\tau)]\exp\left( -S[X]\right)
\label{partf}
\end{equation}
where the action is
\begin{equation}
S[X]=\int d\sigma d\tau \frac{1}{4\pi}\partial_a X(\sigma,\tau)
\cdot \partial_a X(\sigma,\tau)
+\int_0^{2\pi}\frac{d\tau}{2\pi} T(X(\tau)) 
\label{actionint}
\end{equation}
Here, the first term in (\ref{actionint}) is the bulk action and is integrated over the
volume of the unit disc.  The second term in (\ref{actionint})
is integrated on the circle which is the boundary of the unit disc and
describes the interactions.
The scalar fields $X^j$ have $D$ components with $j=1,...,D$ and
$D=26$ for a critical string. In some of the following computations we
will take $D$ as arbitrary and large, but always for applications to
string theory we should eventually put $D$ to  
26.  A $1/D$-expansion generally has a finite radius of convergence
and we expect it to be quite accurate when $D=26$.  We are working in
a system of units where $\alpha'=1$.

By classical power-counting the tachyon field has dimension one and is
a relevant operator. When it is the only interaction, the field theory
is perturbatively super-renormalizable and all ultraviolet divergences
can be removed by normal ordering.  At this point we must distinguish
between the case where $T(X)$ is a polynomial and a non-polynomial
function of $X$.  In the latter case,  it can have large anomalous
dimensions and generally requires a non-perturbative  renormalization
beyond normal ordering. It is well known that this renormlization
makes the beta-function non-linear in $T$, so that if vanishing of the
beta function is taken as the field equation for $T$, these nonlinear
terms describe tachyon
scattering\cite{Klebanov:1987wx,Kostelecky:1999mu}.   This
renormalization would also generate higher derivative counter-terms
and thereby couple all of the other open string degrees of freedom. In
this Paper we will discuss only the case of polynomial potentials with
$O(D)$ symmetry.  We find  that, in the case of non-polynomial
potentials, because of a non-commutativity of the large $D$ and large
cutoff limits, the large anomalous dimensions interfere with the $1/D$
expansion. The analysis with polynomial potentials is still sufficient
to deduce higher derivative terms in the tachyon effective action.

When $T(X)$ and the other fields are adjusted so that the sigma model
that they define is at an infrared fixed point of the renormalization
group, these background fields are a solution of the classical
equation of motion of string theory.   Witten and
Shatashvili~\cite{Witten:1992qy,Shatashvili:1993kk} have argued that
these equations of motion can be derived from an action which is
derived from the disc partition function (\ref{partf}) by a
prescription which we shall make use of below.

We begin with the observation in ref.\cite{Callan:nz} that the
bulk excitations can be integrated out of (\ref{partf}) to get an
effective non-local field theory which lives on the boundary.  To do
this we write the field in the bulk as
$$
X=X_{\rm cl}+X_{\rm qu}
$$
where $$-\partial^2X_{\rm cl}=0$$ and $X_{\rm cl}$ approaches the
fixed (for now) boundary value of $X$,
$$X_{\rm cl}\to X_{\rm bdry}~{\rm and}~X_{\rm qu}\to 0$$ Then, in the
bulk, the functional measure is $dX=dX_{\rm qu}$ and
\begin{equation}
S=\int \frac{d^2\sigma}{4\pi}\partial X_{\rm qu}\cdot\partial X_{\rm
qu} + \int \frac{d\tau}{2\pi} \left\{ \frac{1}{2}X^j |i\partial_\tau|X^j
+T(X) \right\}
\label{1daction}
\end{equation}
Then, the integration of $X_{\rm qu}$ produces a multiplicative
constant in the partition function - the partition function of the
Dirichlet string, which we shall denote $Z_{dir}$.  The kinetic term
in the boundary action is non-local.  The absolute value of the
derivative operator is defined by the fourier transform,
$$
|i\partial|\delta(\tau-\tau')=\sum_n\frac{|n|}{2\pi}e^{in(\tau-\tau')}
$$
The partition function of the boundary theory is then
\begin{equation}
Z=Z_{dir}\int [dX_j]e^{-\int_0^{2\pi}\frac{d\tau}{2\pi}
\left(\frac{1}{2}X^j |i\partial| X^j + T(X)-J\cdot X\right)}
\end{equation}
where we have added a source $J^i(\tau)$ so that the path integral can
be used as a generating functional for correlators of the fields $X_j$
restricted to the boundary.  In particular, this source will allow us
to compute the correlation functions of vertex operators of open
string degrees of freedom.  The remaining path integral over the
boundary $X^j(\tau)$ defines a one-dimensional field theory with
nonlocal kinetic term.  If the tachyon field were absent ($T=0$), the
further integration over $X^j(\tau)$ would give a factor which
converts the Dirichlet string partition function to the Neumann string
partition function.

\vskip .1in
\noindent
{\bf Large $D$ limit}
\vskip .1in

It is interesting to consider the boundary conformal field theory in
the large $D$ limit.  To do this we must assume that the tachyon is a
spherically symmetric function invariant under $O(D)$ rotations and has
the form
$$
T(X(\tau)) = D T\left( \frac{X^2(\tau)}{D}\right)
$$
Since a typical configuration of $X^2(\tau)$ is of order $D$, a
function of this form is itself of order $D$.  In order to study a
large $D$ expansion it is convenient to introduce auxiliary fields.
This is done by inserting unity into the path integral
\begin{equation}
1=\int [d\chi] \delta\left(\chi-\frac{X^2}{D}\right)=\int [d\chi d\lambda]
\exp\left(i\int_0^{2\pi} \frac{d\tau}{2\pi}(D\lambda\chi -\lambda
X^2)\right)
\label{const}
\end{equation}

Here and later in this paper, we shall make use of zeta-function
regularization.  The Riemann zeta function is
$$\zeta(s)=\sum_1^\infty \frac{1}{n^s}$$ 
It converges for $s>1$ and has a
finite analytic continuation for all real $s\neq 1$ where it has a
simple pole.  It has values
$$\zeta(0)=-1/2~~,~~\zeta'(0)=-\frac{1}{2}\ln(2\pi)~~,
~~\zeta(2)=\frac{\pi^2}{6}~~,~~
\zeta(3)=1.20206...$$ Using zeta-function regularization, the Jacobian
wich results from scaling the functional integration variable in
(\ref{const}) by $D$ is
$$ \det\left(D\cdot\delta(\tau-\tau')\right)=
D^{\sum_{n=-\infty}^\infty 1} =D^{1+2\zeta(0)}=1$$

Inserting (\ref{const}) into the functional integral results in
\begin{equation}
Z=Z_{dir}\int [dX d\chi d\lambda] e^{- \int_0^{2\pi}\frac{d\tau}{2\pi}
\left(\frac{1}{2} X^j(|i\partial|+2i\lambda)X^j+ DT(\chi)
-Di\lambda\chi-J\cdot X\right) }
\end{equation}

The variables $X^j(\tau)$ now appear quadratically and can be
integrated out of the partition function.  It will be convenient to do
this only for the non-zero modes of $X^j$.  This will leave behind an
explicit integral over the zero mode which is defined by
$$
\hat X^j = \int_0^{2\pi}\frac{d\tau}{2\pi} X^j(\tau)
$$
The non-zero modes are integrated out to produce the partition function
\begin{equation}
Z=Z_{\rm dir}\int d\hat X^j [d\chi d\lambda] e^{-S_{eff}[\hat
X,\chi,\lambda]}
\label{effpart}
\end{equation}
where
\begin{eqnarray}
S_{\rm eff}= \frac{D}{2}{\rm Tr}\ln\left( {\cal P}\left(
|i\partial|+2i\lambda\right){\cal P}\right)+D\int_0^{2\pi}
\frac{d\tau}{2\pi}\left[ T(\chi) -i\lambda\left(\chi-\frac{\hat
X^2}{D}\right)-\frac{\hat X}{D}\cdot J \nonumber\right. \\
\left.-\frac{1}{2D} \int_0^{2\pi}
\frac{d\tau'}{2\pi}(J(\tau)-2i\lambda(\tau)\hat X)(\tau| {\cal
P}\frac{1}{|i\partial| +2i\lambda}{\cal
P}|\tau')(J(\tau')-2i\lambda(\tau')\hat X)\right]
\label{effact}
\end{eqnarray}
where ${\cal P}$ is the projection operator onto non-zero modes, for
example,
$$
{\cal P}J(\tau)=J(\tau)-\int_0^{2\pi}\frac{d\tau'}{2\pi} J(\tau')
$$
The effective action (\ref{effact}) is divergent and has to be
renormalized.  We will postpone discussion of renormalization
momentarily.

In the large $D$ limit, the functional integrals over $\chi$ and
$\lambda$ in (\ref{effpart}) are evaluated by using the saddle point
approximation.  The saddle point equations obtained by varying the
effective action by $\chi$ and $\lambda$ are
\begin{equation}
T'(\chi)=i\lambda
\label{lambda}
\end{equation}
and
\begin{equation}
\chi(\tau)=\frac{\left(\hat X+x(\tau)\right)^2}{D}+(\tau|{\cal
P}\frac{1}{|i\partial| +2T'(\chi)}{\cal P}|\tau)
\label{chi}
\end{equation}
respectively.  We have used (\ref{lambda}) to eliminate $\lambda$ from
eqn.(\ref{chi}).  $x(\tau)$ is the induced classical field,
\begin{equation}
x(\tau)=\int_0^{2\pi} \frac{d\tau'}{2\pi} (\tau|{\cal P}
\frac{1}{|i\partial| +2T'(\chi)}{\cal P}|\tau')\left(
J(\tau')-2T'(\chi(\tau'))\hat X\right)
\label{classicalfield}
\end{equation}
When eqn.(\ref{chi}) is solved, $\chi$ is a functional of the source
$J$ and the tachyon field $T$.  There is a stability requirement that
the solution is at a minimum of the effective action.

To the leading order in the large $D$ limit, the partition function is
$$
Z=Z_{\rm dir}\int d\hat X e^{-S_{eff}[\chi_0,\lambda_0,\hat X]}
$$
where $\chi_0$ and $\lambda_0$ are the solutions of equations (\ref{chi})
and (\ref{lambda}).   

\vskip .1in
\noindent
{\bf Renormalization}
\vskip .1in

The trace of the logarithm in (\ref{effact}) has ultraviolet
divergences that we must renormalize.  The divergent parts are 
\be
\frac{1}{2}{\rm Tr}\ln\left({\cal P}( |i\partial|+2T'(\chi){\cal
P}\right)= \sum_1^\infty \ln(n) + \int_0^{2\pi} \frac{d\tau}{2\pi}
\left(2T'(\chi(\tau)) \sum_1^\infty
\frac{1}{n}+F(2T'(\chi(\tau)))\right)
\label{tracelog}
\ee where the finite remainder after subtraction of the divergent
terms is \be F(2T'(\chi))=-\int_0^{2\pi}
\frac{d\tau}{2\pi}\int_0^{2\pi} \frac{d\tau'}{2\pi} G(\tau-\tau';0) 2
T'(\chi(\tau))G(\tau-\tau';0) T'(\chi(\tau'))+\dots
\label{finitepart}
\ee and the Green function \be G(\tau-\tau';y(\tau))=(\tau|{\cal
P}\frac{1}{|i\partial|+ y(\tau)}{\cal P}|\tau') \ee The first of the
divergent terms can be handled by zeta-function regularization, where
$$
\sum_1^\infty \ln(n)= -\frac{d}{ds}\zeta(s)|_{s\to 0}
=\frac{1}{2}\ln(2\pi)
$$
The second one is truly divergent and must be subtracted.

Counterterms must be introduced to cancel this divergence.  This is
achieved by the renormalization transform
$$
T(\chi) \to :T\left(\chi-2 \sum_1^\infty\frac{1}{n}-2c_1 \right):$$
where $c_1$ is an arbitrary finite constant. When we substitute this
into the effective action and translate the variable $\chi$, up to un
irrelevant constant it has the form
\begin{eqnarray}
&&S_{\rm eff}=D\left\{\int^{2\pi}_0\frac{d\tau}{2\pi}
\left(:T(\chi):-(\chi-\frac{\hat
X^2}{D}+2c_1):T'(\chi):+F(2:T'(\chi):) \right.\right.  \nonumber \\&&
\left.\left.  -\frac{1}{2D}(J-2:T'(\chi):\hat X)
\frac{1}{|i\partial|+2:T'(\chi):}(J-2:T'(\chi):\hat X)- \frac{\hat
X}{D}\cdot J\right)\right\}~~~~~
\end{eqnarray}
The appearance of the arbitrary constant $c_1$ in the action reflects
the arbitrariness involved in subtracting the divergent terms.  This
arbitrariness was discussed in ref.\cite{Tseytlin:2000mt}.  The
constant $c_1$ should eventually be fixed by some renormalization
prescription.

The replacement of the tachyon field $T$ by $:T:$ amounts to the large
$D$ limit of normal ordering.  We can introduce an ultraviolet cutoff
$\Lambda$ and renormalization scale $\mu$ by the notation \be
\sum_1^\infty \frac{1}{n} = \ln\frac{\Lambda}{\mu}
\label{zeta1}
\ee Then, taking a logarithmic derivative of $:T:$ by $\mu$ leads to a
simple linear beta function for the tachyon field at this order
\cite{Klebanov:1987wx,Kostelecky:1999mu}
$$
\beta(T)= -:T:-2:T:'
$$
which, we shall show in the following is just the large $D$ limit of
the tachyon wave operator.  In the following, we will assume that this
renormalization procedure has be done and will drop the normal
ordering dots from $T$.  The net effect has been to replace the
divergent term linear in $T'$ in the trace-log term in the action by
$T'$ times a finite arbitrary constant.  Now, the effective action and
the eqn.(\ref{chi}) which determines its minimum are free of infrared
divergences.

\vskip .1in
\noindent
{\bf Small derivative expansion}
\vskip .1in

A transparent way to understand the content of the classical
partition function is to consider the limit where $T(X)$ is a smooth
function and to expand in derivatives of $T$.  To do this, we set the
source $J$ to zero.  Then, we expect that the condensate $\chi$ is a
constant, independent of $\tau$.  Then, the Green function can easily
be evaluated.  It is most useful to consider an expansion of
(\ref{chi}) (after renormalization) 
\bea \chi&=& \frac{\hat X^2}{D} -2
c_1+2\sum_{p=1}^\infty\zeta(p+1)\left(-2T'(\chi)\right)^p\cr &=&
\frac{\hat X^2}{D} -2\left( c_1+\gamma+\psi(2T'+1)\right) 
\eea 
where
$\gamma$ is the Euler-Mascheroni constant,  $\psi(x)=d\ln\Gamma(x)/dx$
is the Psi (Digamma) function.  The terms on the right-hand-side of
this equation have increasing numbers of derivatives of $T$.  We can
easily solve it iteratively to obtain $\chi$ to any order in an
expansion in the number of derivatives of $T$ that is desired.  For
example, to order three we have:
\begin{equation}
\chi=\frac{\hat X^2}{D}-2 c_1 -4\zeta(2)T'\left(
\frac{\hat{X}^2}{D}-2c_1\right)+8\zeta(3)\left(T'\left(\frac{\hat{X}^2}{D}
-2c_1\right)\right)^2 +\ldots
\end{equation}
This can then be plugged into equation (\ref{effact}) to get
\begin{equation}
Z=Z_{\rm dir}\int d\hat X_j e^{-DT\left(\frac{\hat X^2}{D}\right)}
\left(1-2 c_1DT'\left(\hat
\frac{X^2}{D}\right) +2D\zeta(2)\left[T'\left(\frac{\hat X^2}{D}\right)\right]^2 
+\ldots\right)
\label{derexp}
\end{equation}
where the omitted terms denoted by dots are at least for orders in
derivatives of $T$ by its argument.

The Witten-Shatashvili action is given by
$$
S=\left( 1+\int \beta(T)\frac{\delta}{\delta T}\right)Z
$$
$$
=Z_{\rm dir}\int d\hat Xe^{-DT\left(\frac{\hat
X^2}{D}\right)} \left\{1+DT\left(\hat \frac{X^2}{D}\right)
+2DT'\left(\hat \frac{X^2}{D}\right)\left[1-
c_1D T\left(\hat \frac{X^2}{D}\right)\right]\right\}
$$
It is not difficult to show that this action exactly coincides
with the one found in~\cite{Tseytlin:2000mt},
where the ambiguity $c_1$ was first discussed. In fact, rescaling
$T$ as $DT(\hat X^2/D)=T(\hat X)$ and rewriting the derivative  with respect
to the argument in terms of derivatives with respect to $\hat X^\mu$, 
$S$ becomes
\be
S=\int d\hat X e^{-T}\left(1+T+(1+c_1)\partial_\mu T\partial^\mu T-
c_1 T\partial_\mu T\partial^\mu T+ O((\partial T)^2)\right)
\ee

\vskip .1in
\noindent
{\bf Higher orders in $1/D$}
\vskip .1in

To compute higher orders in $1/D$, we must consider fluctuations of the
consensate $\chi$.  For this computation, in this section we will set
the source $J$ equal to zero, so that the value of $\chi$ in the
leading order is a constant which satisfies the equation (\ref{chi}).
Thus, we write
$$
\chi(\tau)=\chi_0+\delta\chi
$$
$$
i\lambda=T'(\chi_0)+i\delta\lambda
$$
$\delta\chi$ and $\delta \lambda$ are the quantum variable that have
to be integrated.  $\chi_0$ satisfies
\begin{equation}
\chi_0=\frac{\hat X^2}{D}+(\tau|\frac{1}{|i\partial|
+2T'(\chi_0)}|\tau)'
\label{chi0}
\end{equation}
and, being a constant, can be written explicitly as 
\be
\chi_0=\frac{\hat X^2}{D}+2\ln\frac{\Lambda}{\mu}+ 2 H[2 T'(\chi_0)]
\ee 
where we used (\ref{zeta1}), 
\be 
H(x)=\frac{\partial F}{\partial
x}= \sum_{k=1}^{\infty}(-1)^k x^{k}\zeta(k+1) =-\gamma-\psi(x+1)
\label{H}
\ee 
and $F(2T'(\chi_0))$
is (\ref{finitepart}) the finite part appearing in (\ref{tracelog})
for a constant $\chi$ 
\be 
F(x)=\sum_{k=1}^{\infty}\frac{(-1)^k
x^{(k+1)}\zeta(k+1)}{k+1} =-\gamma x-\ln\Gamma(x+1)
\label{F}
\ee 
Then in the partition function
(\ref{effpart}) the integral over $\chi$ is computed in the large $D$
limit by the saddle point approximation.  It is given by the leading
order 
\bea 
\frac{Z}{Z_{\rm dir}}=\int d\hat X_j \exp\left\{-D\left[
T(\chi_0) - H(2 T'(\chi_0))2 T'(\chi_0)+F(2T'(\chi_0))\right]\right\}\cr
=\int d\hat X_j \exp\left\{-D\left[
T(\chi_0)+2 T'(\chi_0)\psi(2 T'(\chi_0)+1)-\ln\Gamma(2 T'(\chi_0)+1)
\right]\right\}
\eea
multiplied by the determinant of the operator
$$
\left( \matrix{2P(\tau-\tau')+4\frac{\hat X^2}{D}G(\tau-\tau') & -i
\cr -i & T''(\chi_0) \cr }\right)
$$
where \be
P(\tau-\tau';2T'(\chi_0))\equiv(G(\tau-\tau';2T'(\chi_0)))^2=
\left((\tau |{\cal P}\frac{1}{|i\partial|+2T'(\chi_0)}{\cal
P}|\tau')\right)^2
\label{chichi}
\ee 
is the bubble diagram which contributes to the $\chi$-$\chi$
correlator.  Eqs. (\ref{zeta1}) and (\ref{chichi}) can be used to
estimate the divergent part of the $1/D$ terms 
\bea 
&&\frac{1}{2}{\rm
Tr}\ln \left[1+2 T'' \left(P+2\frac{\hat X^2}{D}G\right)\right]=
\frac{1}{2}{\rm \hat Tr}\ln \left[1+2 T''(\chi_0) \left(P+2\frac{\hat
X^2}{D}G\right)\right]\cr&&+ 4T''(\chi_0)\left[\ln\frac{\Lambda}{\mu}+
H(2T'(\chi_0))\right]\left[\ln\frac{\Lambda}{\mu}+
H(2T'(\chi_0))+\frac{\hat X^2}{D}\right] 
\eea 
where the first term is
finite and contains higher powers of $T''(\chi_0)$.  
These terms do not participate to the renormalization program.  To
the order $1/D$ the partition function then reads
\begin{eqnarray}
&&\frac{Z}{Z_{\rm dir}}=\int d\hat X_j \exp\left\{-D\left[ T(\chi_0) -
H(2 T'(\chi_0))2 T'(\chi_0)+F(2T'(\chi_0)) \right.\right.\cr
&&\left.\left.+\frac{4}{D}T''(\chi_0)\left[\ln\frac{\Lambda}{\mu}+
H(2T'(\chi_0))\right]\left[\ln\frac{\Lambda}{\mu}+
H(2T'(\chi_0))+\frac{\hat X^2}{D}\right] \right.\right.\cr
&&\left.\left.+\frac{1}{2D}{\rm \hat Tr}
\ln\left[1+2T''(\chi_0)\left(P+2 \frac{\hat
X^2}{D}G\right)\right]\right] \right\}
\label{effpart2}
\end{eqnarray}

The renormalization of this partition function can be achieved by
normal ordering. Up to order $1/D$ 
\bea
&&:T(\chi):=\exp\left\{\left(\ln\frac{\Lambda}{\mu}+c_1\right)\nabla^2\right\}
T(\chi) =T(\chi+2\ln\frac{\Lambda}{\mu}+2c_1)\cr &&+
\frac{4}{D}\left(\ln\frac{\Lambda}{\mu}+c_1\right)\left(\chi+\ln\frac{\Lambda}{\mu}+c_1\right)T''(\chi+2\ln\frac{\Lambda}{\mu}+2c_1)
\eea 
It is simple to see that in terms of the normal ordered
$:T(\chi):$ the effective action is finite and reads 
\bea &&S_{\rm
eff}=D\left\{T(\chi_0) - H(2 T'(\chi_0))2
T'(\chi_0)+F(2T'(\chi_0))\right.\cr&&\left.  -\frac{4}{D}
T''(\chi_0)\left(c^2_1+c_1\chi_0- \frac{\hat X^2}{D} H(2
T'(\chi_0))-(H(2 T'(\chi_0)))^2\right)\right.\cr
&&\left. +\frac{1}{2D}{\rm \hat Tr} \ln\left[1+2T''(\chi_0)\left(P+2
\frac{\hat X^2}{D}G\right)\right]\right\} 
\label{seffective}
\eea 
where now 
\be
\chi_0=\frac{\hat X^2}{D}-2 c_1+ 2 H[2 T'(\chi_0)] 
\ee 
Moreover any
power of $T''$ in the expansion of ${\rm \hat Tr}\ln[1+2T''(P+2 G \hat
X^2/D )]$ in (\ref{effpart2}) can be computed exactly.
\vskip .1in
\noindent
{\bf The limit of quadratic tachyon profile}
\vskip .1in

As a check of our calculation we take the spherically symmetric 
tachyon profile considered in~\cite{Witten:1992qy}
$$
T(\chi) =\frac{a}{D}+u\chi
$$
For this potential only the leading term in the $1/D$ expansion 
survives (higher order terms contain derivatives with respect to $\chi$
of order 2 and higher and vanish) and from (\ref{H},\ref{F}) and 
(\ref{effpart2}) one gets
\be
\frac{Z}{Z_{\rm dir}}=e^{-a+2Du(c_1+\gamma)}(2u)^{D/2}
\left[\Gamma(2u)\right]^D
\ee
namely Witten's result with the ambiguity due to renormalization 
kept into account.

\vskip .1in
\noindent
{\bf Why the beta function is linear}
\vskip .1in

We shall now perform the calculation of the effective potential in powers
of the tachyon field $T$ and in the large $D$ expansion.
We shall show that these two expansions do not commute and that the
$1/D$ expansion fails in reproducing the large anomalous dimension
\cite{Klebanov:1987wx,Kostelecky:1999mu} 
of non-polynomial tachyon profiles.
 
To compare our approach to that of Klebanov and Susskind 
\cite{Klebanov:1987wx} let us
consider the partition function of the boundary theory in the presence 
of a constant source term of the form $J(k)=-i k$.
\begin{equation}
\frac{Z(k)}{Z_{dir}}=\int [dX_j]e^{-\int_0^{2\pi}\frac{d\tau}{2\pi}
\left[\frac{1}{2}X^j |i\partial| X^j + T(X)\right]-i k\cdot \hat X}
\label{zetak}
\end{equation}
and expand the exponetial in powers of $T(X)$. 
The first non-trivial term is
\be
\frac{Z^{(1)}(k)}{Z_{dir}}=-\int [dX_j]\int \frac{dk_1}{(2\pi)^D}
\int_0^{2\pi}\frac{d\tau_1}{2\pi}T(k_1)e^{-\int_0^{2\pi}\frac{d\tau}{2\pi}
\left(\frac{1}{2}X^j |i\partial| X^j\right)-i k \hat X+ik_1 X(\tau_1)}
\ee
The functional integral over the non-zero modes of $X(\tau)$ gives
\be
\frac{Z^{(1)}(k)}{Z_{dir}}=-\int d\hat X_j\int \frac{dk_1}{(2\pi)^D}
T(k_1)e^{-\frac{k_1^2}{2} G(0)
+i (k_1-k)\hat X}
\ee
the integrals over the zero-modes give a $D$-dimensional $\delta$-function
so that 
\be
-\frac{Z^{(1)}(k)}{Z_{dir}}\equiv T_R(k)=T(k)
e^{-\frac{k^2}{2}G(0) }
\ee
This equation provides the renormalized coupling $T_R$ 
in terms of the bare coupling $T$, to the 
lowest order in perturbation theory. It
corresponds to normal ordering\footnote{To make the
comparison with the results of ref.\cite{Klebanov:1987wx} 
more straightforward, in this section we choose
the renormalization constant ambiguity $c_1=0$.}.

In order to study the large $D$ limit, 
when $T(X)$ is a spherically symmetric function, $T(X)=DT(X^2/D)$,
it is useful to introduce the
Fourier transform  of $T(k)$ in the form
\be
{T}(k)=\int d\hat X e^{-ik\hat X}
\int d\rho DT(\rho)e^{-i\rho \frac{\hat{X}^2}{D}}
=\int d\rho \left(\frac{\pi D}{i\rho}\right)^{D/2} D {T}(\rho)
e^{i\frac{k^2 D}{4\rho}}
\ee
so that the renormalized coupling as a function of $\rho$ reads
\be
T_R(\rho)=\left(1+\frac{i2\rho G(0)}{D}\right)T(\rho)
\label{tren}
\ee

The second order term in $T$ is given by
\be
\frac{Z^{(2)}(k)}{Z_{dir}}=\int [dX]
\int_0^{2\pi}
\frac{d\tau_1}{4\pi}
\frac{d\tau_2}{2\pi}
\int \frac{dk_1dk_2}{(2\pi)^{2D}} {T}(k_1){T}(k_2)
<e^{ik_1X(\tau_1)}e^{ik_2X(\tau_2)}>e^{-i k\hat{X}}
\ee
The integral over the non-zero modes can be performed to give
\bea
\frac{Z^{(2)}(k)}{Z_{dir}}=
\int_0^{2\pi}
\frac{d\tau_1}{4\pi}
\frac{d\tau_2}{2\pi}
\int \frac{dk_1dk_2}{(2\pi)^{2D}} \int d\hat X \int d\rho_1 d\rho_2 D^2 
T(\rho_1)T(\rho_2)\left(-\frac{\pi^2 D^2}{\rho_1\rho_2}\right)^{D/2}\cr
\exp\left[i(k_1+k_2-k)\hat X +
i\frac{k_1^2 D}{4\rho_1}+i\frac{k_2^2 D}{4\rho_2}+
-\frac{1}{2}\left(k_1^2 +k_2^2\right)G(0)-k_1k_2G(\tau_1 -\tau_2)\right]
\eea
Integrating over $k_1$ and $k_2$ and expanding for large $D$ one gets

\be
\frac{Z^{(2)}(k)}{Z_{dir}}=
\int_0^{2\pi}
\frac{d\tau_1}{4\pi}
\frac{d\tau_2}{2\pi}
\int d\hat X
\int d\rho_1 d\rho_2 D^2 {T_R}(\rho_1) {T_R}(\rho_2) 
e^{-\frac{2\rho_1 \rho_2}{D}G^2(\tau_1 -\tau_2)
-ik\hat X-i(\rho_1 +\rho_2)\frac{\hat X^2}{D}}
\label{zeta2}
\ee
Performing at this step the integral
over $\tau_1$ and $\tau_2$, for $\rho_1 \rho_2 <0$ one finds
a logarithmic divergence which would reproduce the large anomalous
dimension. This divergence
would provide the non-linear part of the beta function for non-polynomial
tachyon fields.
If, instead, in eq. (\ref{zeta2})
one first expands the exponential in powers of $1/D$
and then performs the integral
over $\tau_1$ and $\tau_2$, the result is finite and reads
\be
\frac{Z^{(2)}(k)}{Z_{dir}}=
\int d\hat X
\int d\rho_1 d\rho_2D {T_R}(\rho_1) {T_R}(\rho_2)\left[\frac{D}{2}
-2\zeta(2)\rho_1\rho_2 \right]
e^{-ik\hat X-i(\rho_1 +\rho_2)\hat X^2/D}
\ee
In terms of $T(\hat X^2/D)$, this can be rewritten as
\be
\frac{Z^{(2)}(k)}{Z_{dir}}=
\int d \hat{X}\left[\frac{D^2}{2}\left(T_R({\hat X^2}/{D})\right)^2
+2\zeta(2) D \left(T_R^{'}({\hat X^2}/{D})\right)^2\right]e^{-ik\hat X}
\label{ks}
\ee
Eq.(\ref{ks}) coincides with eq.(\ref{derexp}) when $c_1=0$, 
the exponetial is expanded up to the second order in $T$
and a constant source is introduced.

\vskip .1in
\noindent
{\bf Conclusions}
\vskip .1in
We have studied the problem of tachyon condensation in bosonic open string
theory in the case where the tachyon condensate has spherical symmetry.
Our analysis is limited to the case where the condensate field is a polynomial.
In this case, the beta function is linear.  Furthermore, the only fields which
are polynomial and are fixed points are $T=0$ and $T=\infty$.  These are the
usual perturbative and stable vacua which are found in the special case where the
tachyon is a quadratic function of $X$.  In the present work, we have an elaboration of that case to higher order polynomial potentials.

It would be straightforward to obtain boundary states for D-branes with a tachyon condensate in the large $D$ limit which we have considered here.  When
$D=26$ these would not be exact boundary states, but would be systematically correctable to any order in $1/D$.  It would be interesting to examine whether, off-shell, they
require extra constraints to define them as was discussed for the quadratic tachyon profile in refs.\cite{Laidlaw:2001jt}.

\end{document}